# Direct Evidence of Torque-mediated Optical Magnetism

**M. T**uan **T**rinh[1*], **K**rishnandu **M**akhal[1], **E**lizabeth **F.C. D**reyer[1], **A**poorv **S**hanker[2], **S**eong-**J**un **Y**oon[2], **J**insang **K**im[2], and **S**tephen **C. R**and[1,3]

[1]*Dept. of Electrical Engineering, University of Michigan, Ann Arbor, MI 48109*
[2]*Dept. of Materials Science, University of Michigan, Ann Arbor, MI 48109*
[3]*Dept. of Physics, University of Michigan, Ann Arbor, MI 48109*
*\*tuantrin@umich.edu*

**Abstract:** We report experimental evidence of a mechanism that supports and intensifies induced magnetization at optical frequencies without the intervention of spin-orbit or spin-spin interactions. Energy-resolved spectra of scattered light, recorded at moderate intensities ($10^8$ W/cm$^2$) and short timescales (<150 fs) in a series of non-magnetic molecular liquids, reveal the signature of torque dynamics driven jointly by the electric and magnetic field components of light at the molecular level. While past experiments have recorded radiant magnetization from magneto-electric interactions of this type, no evidence has been provided to date of the inelastic librational features expected in cross-polarized light scattering spectra due to the Lorentz force acting in combination with optical magnetic torque. Here, torque is shown to account for inelastic components in the magnetic scattering spectrum under conditions that produce no such features in electric dipole scattering, in excellent agreement with quantum theoretical predictions.

## 1. Introduction

For several decades, interactions that couple electric and magnetic effects in bulk materials through magnetostriction and piezoelectricity have been the subject of intense investigation. Such interactions operate either by magnetostriction in a material which generates an internal piezo-electric field in response to the deformation or by the reverse process. They are of great potential value for energy conversion and novel sensor technology. Consequently, there is considerable interest in creating multifunctional circuits based on magneto-electric and multiferroic materials that combine electronic and magnetic properties [1-9]. To date, however, such devices have relied on the properties of bulk magnetic solids or magnetic thin films, making them difficult to miniaturize. The control of magnetism in nanoscale or molecular level is of great interest for the advancement of nanotechnology. Yet, there have been no reports of induced magnetism in non-magnetic materials on the nanoscale. While magneto-electric interactions on nanoscale or molecular level have been discussed in the literature [10-12], and recently observed [13] and theorized [14,15], no detailed confirmation of the theoretical torque mechanism proposed to explain how these effects undergo strong enhancement has been published. At non-relativistic intensities of light, possible effects of the optical magnetic field are normally dismissed owing to the smallness of the Lorentz driving force $qvB$ (proportional to the velocity v of a charge $q$) compared to the electric force $qE$. Since $B<<E$, this would be perfectly justified if no enhancement mechanism existed to magnify the Lorentz force of light. However, a fundamental mechanism exists that is capable of enhancing magnetism induced jointly by the optical fields $E$ and $H$ at the level of individual molecules [14, 15]. Under magneto-electric interaction, orbital angular momentum converted to molecular rotational angular momentum enlarging the effective current area. As a result, the magnetic dipole moment enhanced. In this paper, we present the first direct experimental evidence that torque by the optical magnetic field, thought to be capable of enhancing radiant magnetization to the same level as electric dipole polarization at optical frequencies under non-relativistic conditions[13-15], does indeed accompany magneto-electric interactions at the molecular level.

    Here, ultrafast magnetic response was studied in pulsed optical experiments using a simple 90° scattering geometry (Figure 1(a)). Despite the extensive literature on depolarized light scattering [16-22], there have been few reports of the complete radiation patterns or inelastic spectra that are necessary to distinguish between electric dipole (ED) and magnetic dipole (MD) response at the molecular level. Early studies of collisional effects in gases and liquids showed that they generated a small amount of depolarization even in isotropic molecules at low light intensities, on relatively long timescales (i.e. timescales exceeding the collisional reorientation time). However no experiments were performed at high enough intensities or sufficiently short timescales to discriminate between collisional depolarization and magnetization on the basis of the angular distribution of scattered radiation or by performing suitable spectral analysis. In the present work, we utilized moderately high intensities (I~$10^8$W/cm$^2$) and ultrashort pulses ($\tau_p$<*150 fs*) to record both the radiation patterns and spectrally-resolved inelastic feature in polarization-analyzed scattered light.

Our experimental results disclose key details of the mechanism governing radiant optical magnetization in dielectric media for the first time.

The magnitude of an induced magnetic dipole moment *m* is normally limited to a small fraction of the induced electric dipole *p*, namely $(m/p) < \alpha$, where $\alpha=1/137$ is the fine structure constant. Exceptions occur in media where spin-spin interactions are strong or the ED approximation is not upheld, such as in ferromagnets, structured dielectrics, metamaterials, and nanoparticles. However strong induced magnetic response has not been anticipated in natural homogeneous (non-magnetic) materials at high frequencies. For this very reason, highly engineered, non-uniform metamaterials have attracted much attention as an important way to realize magnetic response through structural design. In the present work, however, the mechanism of an intriguing alternative is explored. Dynamic magneto-electric interactions are tested as a route for superseding traditional limitations on induced magnetic moments, and are found to agree with classical simulations [14] and quantum theory [15] predicting that magnetic optical torque can create and enhance magnetic response in nominally non-magnetic media through the ultrafast exchange of orbital and rotational angular momenta.

## 2. Methods and Results

Our experimental setup utilized an amplified femtosecond laser system (Amplitude Inc.) operating at 800 nm and delivering pulses of 0.5 mJ at a rate of 10 kHz over an electronically-tunable bandwidth of 15 to 100 nm. The laser output was used to probe the induced magnetic response of dielectric liquids composed of tetrahedral molecules. Samples included $CCl_4$, $SiCl_4$, $SiBr_4$, $Si(OCH_3)_4$, and $Si(OC_2H_5)_4$. These molecules exhibit isotropic optical response at low powers and represent a series in which the moment of inertia increases systematically. Complete radiation patterns were recorded in co-polarized and cross-polarized light-scattering geometries at 90° with respect to the incident beam by rotating the input polarization angle $\theta$ through 360 degrees under computer control [13]. A schematic of the setup and a typical set of raw data for co- and cross-polarized signals for $CCl_4$ are shown in Figures 1(a) and 1(b). The corresponding radiation patterns are plotted in Figure 1(c), after subtraction of the depolarized components (constant backgrounds) shown in Figure 1(b). The dependence on input intensity of the scattering light was previously shown to be quadratic at low intensities [13]. Radiation patterns for all samples showed the same depolarized and dipolar components with low residuals as in $CCl_4$.

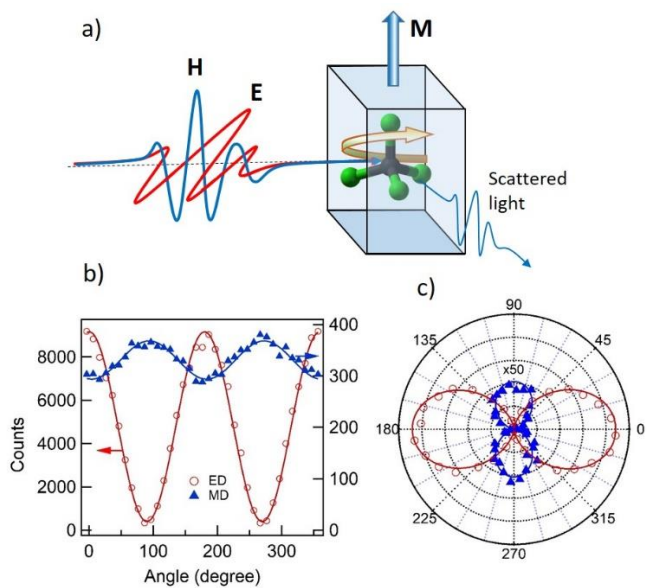

Fig. 1. (a) The light scattering process: the optical electric field causes an electric dipole transition which imparts orbital kinetic energy to the molecule. This energy is subsequently converted to librational motion by the optical magnetic field. The induced electric and magnetic dipole moments cause incoherent scattering at an angle to the incident beam. (b) Raw data for co-polarized (circles) and cross-polarized (triangles) scattering intensities versus input rotation angle in $CCl_4$. (c) Radiation patterns of the co- and cross-polarized light-scattering data after subtraction of the constant (unpolarized) background evident in Fig. 1(b). The solid curves are fits to a $\cos^2\theta$ function.

A key result of this work is shown in Figure 2, which displays the normalized co-polarized and cross-polarized spectra of scattered light in $CCl_4$ recorded with a 0.5 m grating spectrometer, labelled ED and MD, respectively. The co-polarized (Rayleigh scattering) spectrum in red is virtually indistinguishable from the instrumental response over the bandwidth reflecting the pulse duration, the inset of figure 2. The cross-polarized spectrum in blue has large additional features in it. These have been emphasized by subtracting the co-polarized signal to obtain the difference spectrum shown in grey. The extra features appearing in the cross-polarized spectrum are then found to correspond to inelastic scattering from known rotations and vibrations of individual $CCl_4$ molecules. Below the grey curve are the expected positions and relative heights of rotationally- and vibrationally-shifted satellite lines, indicated by vertical bars displaced from the origin (downward arrow at the peak of the pulse spectrum). Co- and cross-polarized spectra for the other samples are shown in Figure 3 (a)-(c). Experimental values of the rotation and vibration energies were determined from a best fit convolution of the instrumental response $I(\omega)$ with an assumed spectrum of satellite lines of variable height and position. The widths of these features was assumed to be instrument-limited. The results for fitted vibration and rotation frequencies in Figure 2 are in good agreement with literature values [24,25]. The comparison of experimental rotation frequencies with results from prior spectroscopy shown in Figure 3(d) provides compelling evidence for our rotational assignments and confirms that rotations (or more correctly librations) are generated during the interaction responsible for cross-polarized scattering.

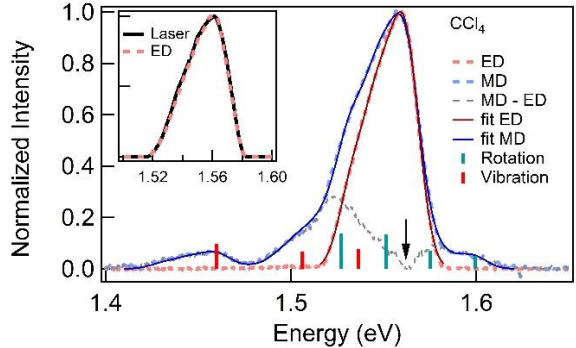

Fig. 2. Normalized co- and cross-polarized scattered light spectra in $CCl_4$, dashed-red (labelled ED) and -blue (labelled MD), respectively. The black arrow indicates the spectral peak. The solid curve showing a best fit that takes instrumental linewidth into account together with inelastic components due to vibrational and rotational transitions as indicated by vertical red and blue bars. The grey curve is the difference between the red and blue curves, highlighting the inelastic components in the MD spectrum. Inset: A comparison of the laser and ED spectra.

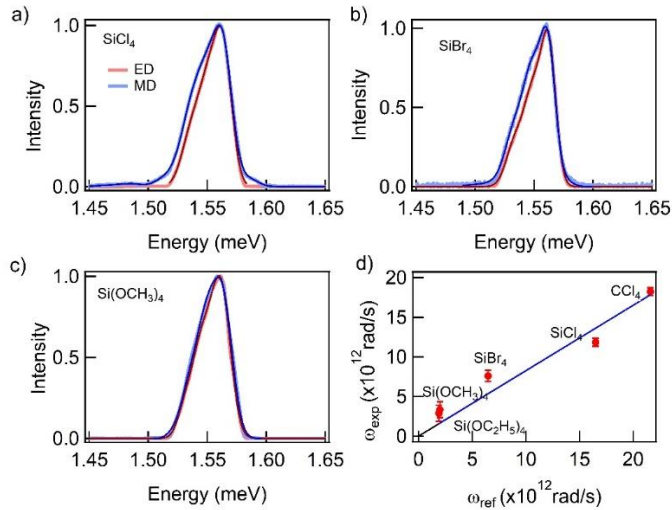

Fig. 3. Normalized co- and cross-polarized scattered light spectra for various compounds. (a) $SiCl_4$, (b) $SiBr_4$ and c) $Si(OCH_3)_4$ with fitted curves. (d) The experimental rotation frequencies ($\omega_{exp}$) plotted versus literature values ($\omega_{ref}$) [24, 25] for all samples. The solid line is a linear fit with a slope of 0.85.

Further measurements supported the proposed 2-photon picture of the interaction. The results of analyzing polarization states of the inelastic scattered light components are shown in Figure 4 and agree with expectations based on the two downward transitions in Figure 5(a). Complete scattered light spectra were recorded at values of $\theta$ separated by discrete steps of 30º to determine whether particular spectral features were polarized, unpolarized or of mixed polarization. In Figure 4(a), the elastic component at the center of the cross-polarized (MD) spectrum dropped by 30% as the input field was rotated away from horizontal (90°). This result indicated that it was partially polarized, composed of both a dipolar contribution to scattered light and a large unpolarized background. Spectral features at large shifts (1.4 to 1.53 eV) on the other hand did not vary at all as the input polarization was varied. This is shown on the left side of Figure 4(a) where all curves are overlapped. Hence these features were completely unpolarized. In the next section, we discuss in more detail the close correspondence of these findings with expectations from the quantum theory of magneto-electric interactions at the molecular level.

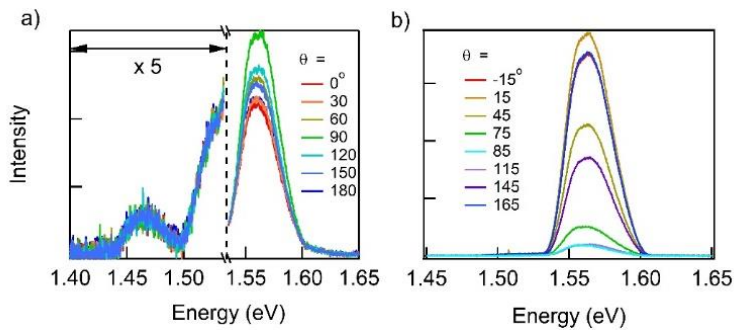

Fig. 4. Polarization dependence of cross-polarized (a) and co-polarized (b) spectra in CCl$_4$. $\theta$ is the pump polarization angle. In (a) the window from 1.40 to 1.53 eV has been magnified 5 times for clarity to show there is no polarization dependence. The input angle $\theta=0°$ corresponds to vertical polarization. In (b) the entire spectrum is highly polarized.

## 3. Discussion

Many physical processes cause depolarization in light scattering. The term itself refers to either loss of polarization or the generation of components orthogonal to the input fields. Hence, depolarization may originate from collisionally-induced rotations, collisional dipole-dipole interactions, electric torque acting on anisotropic electric polarization (in the case of anisotropic molecules), or electric torque acting on Kerr-induced anisotropy in isotropic molecules. Depolarization can also arise from magneto-electric interactions at the molecular level. Of these various processes, only dipole-dipole collisions, Kerr effects, and magneto-electric interactions can occur on timescales faster than the molecular re-orientation time. So these are the only candidates for explaining cross-polarized components in nonlinear light scattering. However, dipole-dipole interactions and Kerr effects yield radiation patterns that are very different from those measured experimentally. Colliding molecules could instantaneously produce a depolarized probe field $E_{probe}$ capable of mediating an optical Kerr polarization [26] of the form $P_z=\varepsilon_0\chi_{zzxx}E_zE_xE_x$, but the angle-averaged polarization $P_z$ would be zero in view of the randomly oriented probe field. This Kerr mechanism therefore does not contribute significantly to the scattering light. At sufficiently high intensities, a Kerr effect of the form $P_x=\varepsilon_0\chi_{xxxx}E_xE_xE_x$ based on a polarized probe input field could in principle contribute importantly to co-polarized light scattering. However, the three fields that drive the nonlinearity would each contribute an angular projection to the radiation pattern. Hence the angular dependence of co-polarized light scattering would vary as $I(\theta)\sim\cos^6\theta$, resulting in a pattern which is not consistent with the data of Figure 1. Hence, in isotropic media such as liquid CCl$_4$ only the magneto-electric mechanism can produce cross-polarized light on femtosecond timescales with both a simple dipole and an unpolarized component as displayed in Figure 1, in the non-relativistic intensity range of our experiments.

Existing theory of magneto-electric interactions at the molecular level [14] suggests that they proceed as shown in Figure 5. The process relies on first establishing an electric polarization in the system with the optical *E* field and then converting the orbital angular momentum of the excited state to rotational angular momentum through a torque interaction driven by *H*. During the magnetic transition the angular momentum axis rotates as described in Refs. 14 and 15, causing an enhancement of the induced magnetic moment. In Figure 5(a), two magnetic transitions are indicated by downward red arrows. These two transitions are driven by frequency components of the incident pulse that are either at the carrier frequency or at a frequency shifted down by the molecular rotation frequency $\omega_\phi$. The former transition (dashed arrow) is detuned from resonance and is therefore expected to be polarization-preserving. The latter transition (solid arrow) resonantly stimulates a rotational excitation which causes depolarization of Stokes-shifted scattering. These two predictions are in excellent agreement with the results in Figures 2 and 4 where the extra spectral features and their polarizations are evident. So while it is true that the radiation patterns of Figure 1 are in complete accord with a magneto-electric interaction at the molecular level, it is more significant in terms of the objectives of the present paper. The results in Figures 2 - 4 agree with the predicted polarizations as well as predicted rotational inelastic features in the scattered light spectra.

We note that the direct stimulation of depolarizing rotations by the resonant magnetic transition in Fig. 5(a) – the solid downward arrow - should cause "knock-on" vibrational excitations too. That is, the energy of a stimulated libration in combination with the energy of translation of a single molecule induces vibrations in neighboring molecules. This is consistent with the presence of Stokes-shifted *vibrational* features in the cross-polarized spectrum of Figure 2. Moreover, the spectral features on the anti-Stokes or high- energy side of the spectrum in Figure 2 are uniquely attributable to the "reverse" transition sequence depicted in Figure 5(b). Beginning from a thermal population in the first excited rotational state $|3\rangle$ at room temperature (labelled with rotational quantum number $J = 1$), a "reverse" magneto-electric transition annihilates a rotational quantum. This process yields an anti-Stokes rotational line but at the same time removes the rotational excitation necessary to cause "knock-on" vibrations during anti-Stokes scattering. Thus, anti-Stokes vibrational features are not expected and indeed no features are observed in Figure 2 at vibrational shifts on the high frequency side of the carrier. On the other hand, a "reverse" transition originating from the second rotational energy level at room temperature could yield a second rotational anti-Stokes line (omitted from the schematic of Figure 5(b) for simplicity). This transition accounts for the second anti-Stokes feature visible in Figure 2, close to 1.6 eV.

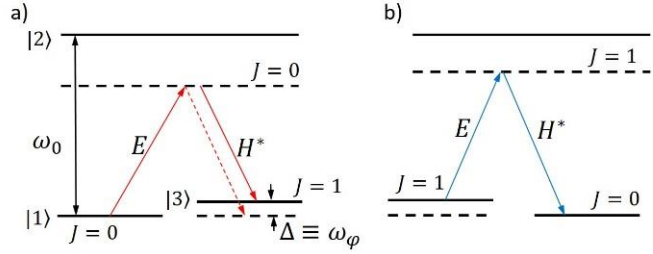

Fig. 5. Two-photon transitions responsible for second-order magneto-electric scattering driven by *E* and *H* fields. (a) In the Stokes process, the *H* field stimulates two classes of magnetic transition back to the ground state, as indicated by the two downward arrows. The magnetic transition at frequency $\omega - \omega_\phi$ (solid downward arrow) is resonant but can only be driven by a Fourier component of the optical pulse that is down-shifted from the carrier frequency by $\omega_\phi$. The first transition (dashed arrow) is ineffective in generating molecular rotations and produces polarized scattering. The second transition stimulates molecular rotations resonantly (final rotational state $J = 1$), giving rise to unpolarized scattering and knock-on vibrations. (b) In the "reverse" process, the *E* field first drives an ED transition which preserves the initial rotational state ($J=1$) at room temperature. Then the magnetic field stimulates an anti-Stokes MD transition at frequency $\omega + \omega_\phi$ which removes the rotation. This generates unpolarized anti-Stokes rotational scattering without the possibility of knock-on vibrations.

The magnetic torque interaction analyzed in Refs. 14 and 15 enables magnetic transitions to take place at the optical frequency as indicated in Figure 5. While the magnetic transitions, labelled by $H^*$ in this figure, satisfy the quantum mechanical selection rules, they must also be ultrafast in order for the magneto-electric interaction to go to completion during the short pulse durations of our experiments. One can check that this requirement is met by turning to the quantum mechanical torque equation to estimate the time scale of magnetic torque dynamics. In the Heisenberg picture, the expectation value of torque expressed in terms of excited state orbital angular momentum *L* is

$$\left\langle \frac{dL_z}{dt} \right\rangle = \frac{i}{\hbar} Tr\{\tilde{\rho}, [H^{(m)}, L_z]\}, \qquad (1)$$

where $\tilde{\rho}$ is the slowly-varying amplitude of the density matrix. Inserting the torque Hamiltonian from Ref. [14]

$$H^{(m)} = (\hbar f L'_- O'_+ a^+ - h.c.), \qquad (2)$$

Eq. (1) can be evaluated for the energy level picture of Figure 5. As shown in the appendix B, the torque equation reduces to

$$\hbar/\tau = \mu_0^{(e)} E, \qquad (3)$$

where $\tau$ is the characteristic time for completion of the dynamics and formation of the enhanced magnetic moment (see appendix B). Transition dipole moments of the tetrahalide series range from $\mu_0^{(e)} = 4.1 \times 10^{-30}$ *C.m* for $CCl_4$ to $\mu_0^{(e)} = 11.3 \times 10^{-30}$ *C.m* for $SiCl_4$ [27]. Consequently torque completion times for the various samples, at an intensity of $I = 10^{10}$ W/cm$^2$ (or $E = 1.94 \times 10^8$ V/m), comparable to our experimental intensities, are found to range from $\tau = 133$ fs down to $\tau = 48$ fs. The implication is that magnetic torque can indeed cause the exchange of orbital angular momentum for rotational angular momentum on an ultrafast timescale, allowing the magneto-electric interaction to go to completion during each pulse.

## 4. Conclusions

In summary, we have observed stimulated molecular rotations in a series of liquids composed of spherical top molecules on which optical electric fields are unable to exert torque, at intensities far below the stimulated rotational Raman threshold. The rotations are therefore uniquely attributable to the effect of the optical magnetic field, as predicted in recent quantum theory of molecular-scale magneto-electric interactions. The experimental evidence supports the conclusion that intense radiant magnetization relies on inter-conversion of orbital and rotational angular momenta driven jointly by electric and magnetic fields. Magneto-electric torque dynamics increase the effective area enclosed by displacement polarization currents in dielectric media, readily yielding enhanced magnetic response as large as electric dipole response [13-15]. Analysis of polarization states of Stokes- and anti-Stokes-shifted librations and vibrations in magnetic dipole spectra in tetrahedral molecules is also fully consistent with the theory of magneto-electric interactions at the molecular level. This unusual nonlinearity may therefore provide a new route for the control of magnetic permeability and dispersion in dielectric media.

Appendix A: **Sample preparation**

The anhydrous chemicals used in this research were spectroscopic grade, purchased from Sigma Aldrich Inc., and were examined in quartz cuvettes with very high quality surfaces after rinsing with DI water, acetone, and isopropanol in a sequence and transferring them immediately to an oven, where they were heated at a temperature > 82 C for ~ 20 minutes. Cleaning and sealing of the cuvettes was performed inside a glovebox to prevent hydration of the samples. Just before each measurement, the four outer surfaces of the sample cuvette were given a final cleaning with optical tissue wetted with MeOH and blown dry using pressurized nitrogen gas.

Appendix B: **Theoretical dynamics for torque completion time**

The time required for the optical magnetic field to rotate the quantization axis of angular momentum in a molecule to rotational angular momentum, thereby enhancing its magnetic moment, was estimated with a classical argument in [14]. Here we provide a quantum derivation based on the torque Hamiltonian derived in that work.
In the Heisenberg picture, the expectation value for the rate of change of the $z$-projection of angular momentum due to an optical magnetic interaction for the basis of states 1, 2, and 3 shown in Figure 5 is

$$\langle \frac{dL_z}{dt} \rangle = \frac{i}{\hbar} Tr\{\tilde{\rho}, [H^{(m)}, L_z]\}, \tag{A.1}$$

where $\tilde{\rho}$ is the slowly varying amplitude of the density matrix. The non-zero terms of the trace are

$$\langle \frac{dL_z}{dt} \rangle = \frac{i}{\hbar} \{\tilde{\rho}_{23}[H^{(m)}, L_z]_{32} + \tilde{\rho}_{32}[H^{(m)}, L_z]_{23}\}, \tag{A.2}$$

and the Hamiltonian for a magnetic dipole interaction including torque [14] is

$$H^{(m)} = \hbar f L'_- O'_+ a^+ + h.c. \tag{A.3}$$

In this expression $f$ is the magnetic interaction strength, and $L'_\mp$, $O'_\pm$ and $a^\pm$ are raising and lowering operators for orbital and rotational angular momenta and the optical field respectively. The expectation of the rate of change of the $z$-projection of angular momentum may now be evaluated as

$$\langle \frac{dL_z}{dt} \rangle = \frac{i}{\hbar} \{-\tilde{\rho}_{23}\langle 3|L_z \hbar f L'_- O'_+ a^+|2\rangle + \tilde{\rho}_{32}\langle 2|\hbar f^* L'_+ O'_- a^- L_z|3\rangle\} = 2i\sqrt{n}\{f\tilde{\rho}_{23} + f\tilde{\rho}_{32}\}. \tag{A.4}$$

The magnetic interaction strength $f$ depends on the electric field per photon $\xi$ and the effective magnetic moment $\mu_{eff} = (\omega_0/\omega_c)\mu_0^{(m)}$, where $\omega_c$ is the cyclotron frequency and $\mu_0^{(m)} = (e\hbar/2m_e)$. The irreducible form of $f$ is [14]

$$\hbar f = i\mu_{eff}\xi/c. \tag{A.5}$$

Once the effective magnetic moment acquires the enhanced value of $\mu_{eff} = c\mu_0^{(e)}/2$ as the result of the molecule exchanging orbital for rotational angular momentum, we find

$$f = i\mu_0^{(e)}\xi/2\hbar = ig/2. \tag{A.6}$$

Assuming the applied optical fields are high enough to generate the maximum value for the off-diagonal element of the density matrix, which is $\tilde{\rho}_{23} = \tilde{\rho}_{32} = \frac{1}{2}$ in a 2-level system, we finally obtain

$$\langle \frac{dL_z}{dt} \rangle = -|g|\sqrt{n}. \tag{A.7}$$

On the left side of the equation above, note that the $z$-projection of orbital angular momentum is only allowed to change by $\hbar$ during a magnetic dipole transition. If the characteristic time for completion of the torque dynamics responsible for enhanced magnetic response is designated by $\tau$, the left side is given by $\langle \frac{dL_z}{dt} \rangle = -\frac{\hbar}{\tau}$ and the equation can be re-written as

$$\frac{\hbar}{\tau} = |g|\sqrt{n}. \tag{A.8}$$

Finally, since $\mu_0^{(e)}E = g\sqrt{n}$, one can solve for the torque completion time to obtain the result

$$\tau = \frac{\hbar}{\mu_0^{(e)}E}. \tag{A.9}$$

As stated in the main text, the timescale on which optical magnetic torque converts orbital angular momentum to rotational energy varies from $\tau = 133\,fs$ in CCl$_4$ to $\tau = 48\,fs$ in SiCl$_4$. The conclusion may therefore be drawn that magnetic torque can mediate the exchange of angular momentum on an ultrafast timescale, allowing a magneto-electric interaction in which both the electric and magnetic field components of the light cause transitions at the optical frequency during a single pulse, molecule by molecule.

**Funding**. MURI Center for Dynamic Magneto-optics, the Air Force Office of Scientific Research (FA9550-12-1-0119 and FA9550-14-1-0040); DURIP grant (FA9550-15-1-0307).

**Acknowledgements.** We thank H. Winful for useful discussions.